\documentstyle[11pt]{article}
 1
\font\elevenrm=cmr10 scaled\magstep 1
 1

\newcommand{\beqn}{\begin{eqnarray}}
\newcommand{\eeqn}{\end{eqnarray}}

\renewenvironment{thebibliography}[1]
 { \elevenrm
   \begin{list}{\arabic{enumi}.}
    {\usecounter{enumi} \setlength{\parsep}{0pt}
     \setlength{\itemsep}{3pt} \settowidth{\labelwidth}{#1.}
     \sloppy
    }}{\end{list}}
\makeatletter
\def\section{\@startsection{section}{0}{\z@}{5.5ex plus .5ex minus
  1.5ex}{2.3ex plus .2ex}{\bf}}
\renewcommand{\theequation}{\rm\thesection.\arabic{equation}}

\newcommand{\appendixA}{\setcounter{equation}{0}
  \def\theequation{\rm{A}.\arabic{equation}}\section*}
\newcommand{\appendixB}{\setcounter{equation}{0}
  \def\theequation{\rm{B}.\arabic{equation}}\section*}
\newcommand{\appendixC}{\setcounter{equation}{0}
  \def\theequation{\rm{C}.\arabic{equation}}\section*}

\newcommand{\sectionA}{\setcounter{equation}{0}
  \def\theequation{\rm{1}.\arabic{equation}}\section*}
\newcommand{\sectionB}{\setcounter{equation}{0}
  \def\theequation{\rm{2}.\arabic{equation}}\section*}
\newcommand{\sectionC}{\setcounter{equation}{0}
  \def\theequation{\rm{3}.\arabic{equation}}\section*}
\newcommand{\sectionD}{\setcounter{equation}{0}
  \def\theequation{\rm{4}.\arabic{equation}}\section*}
\newcommand{\sectionE}{\setcounter{equation}{0}
  \def\theequation{\rm{5}.\arabic{equation}}\section*}
\makeatother

\makeatletter
\newcommand{\fcaption}[1]{
        \refstepcounter{figure}
        \setbox\@tempboxa = \hbox{\small Fig.~\thefigure. #1}
        \ifdim \wd\@tempboxa > 14.4cm
           {\begin{center}
        \parbox{14.4cm}{\small Fig.~\thefigure. #1}
            \end{center}}
        \else
             {\begin{center}
             {\small Fig.~\thefigure. #1}
              \end{center}}
        \fi}

\newcount\@tempcntc
\def\@citex[#1]#2{\if@filesw\immediate\write\@auxout{\string\citation{#2}}\fi
  \@tempcnta\z@\@tempcntb\m@ne\def\@citea{}\@cite{\@for\@citeb:=#2\do
    {\@ifundefined
       {b@\@citeb}{\@citeo\@tempcntb\m@ne\@citea\def\@citea{,}{\bf ?}\@warning
       {Citation `\@citeb' on page \thepage \space undefined}}%
    {\setbox\z@\hbox{\global\@tempcntc0\csname b@\@citeb\endcsname\relax}%
     \ifnum\@tempcntc=\z@ \@citeo\@tempcntb\m@ne
       \@citea\def\@citea{,}\hbox{\csname b@\@citeb\endcsname}%
     \else
      \advance\@tempcntb\@ne
      \ifnum\@tempcntb=\@tempcntc
      \else\advance\@tempcntb\m@ne\@citeo
      \@tempcnta\@tempcntc\@tempcntb\@tempcntc\fi\fi}}\@citeo}{#1}}
\def\@citeo{\ifnum\@tempcnta>\@tempcntb\else\@citea\def\@citea{,}%
  \ifnum\@tempcnta=\@tempcntb\the\@tempcnta\else
   {\advance\@tempcnta\@ne\ifnum\@tempcnta=\@tempcntb \else \def\@citea{--}\fi
    \advance\@tempcnta\m@ne\the\@tempcnta\@citea\the\@tempcntb}\fi\fi}
\makeatother

\def\DIHEP{{\it Davis Institute for High Energy Physics, 
Department of Physics}\\
  {\it University of California, Davis, CA 95616 USA} \\}

\renewcommand{\thefootnote}{\fnsymbol{footnote}}
\begin{document}
\noindent
\thispagestyle{empty}
\begin{flushright}
{\bf UCD-97-16}\\
{\bf hep-ph/9708264}\\
\end{flushright}
  \vspace{0.25cm}
\begin{center}
\begin{bf}
 \begin{Large}
MESSENGER SNEUTRINOS AS COLD DARK MATTER \\
 \end{Large}
\end{bf}
  \vspace{.5cm}
  \begin{large}
  Tao Han and
  Ralf Hempfling\\
  \end{large}
  \vspace{0.5cm}
  \DIHEP
  \vspace{3.5cm}
  {\bf Abstract}\\
\vspace{0.5cm}
\noindent
\begin{minipage}{13.0cm}
\begin{small}
In models where supersymmetry breaking is communicated
into the visible sector via gauge interactions
the lightest supersymmetric particle is typically
the gravitino
which is too light to account for cold dark matter.
We point out that the lightest messenger sneutrinos
with mass in the range of one to three TeV
may serve as cold dark matter over most of
the parameter space due to one-loop electroweak radiative corrections.
However, in the minimal model this mass range has been
excluded by the direct dark matter searches.
We propose a solution to this problem by introducing terms
that explicitly violate the messenger number.
This results in low detection rate for both direct and indirect
searches and allows messenger sneutrinos to be a valid
dark matter candidate in a wide region of SUSY parameter space.

\end{small}
\end{minipage}
\end{center}
\vfill
\vspace{0.25cm}
\vbox{
      \hbox{July, 1997}
      \hbox{\mbox{}}}
\setcounter{footnote}{0}
\renewcommand{\thefootnote}{\alph{footnote}}
\newpage
\vspace{1.2cm}

\def\spose#1{\hbox to 0pt{#1\hss}}
\def\lsim{\mathrel{\spose{\lower 3pt\hbox{$\mathchar"218$}}
     \raise 2.0pt\hbox{$\mathchar"13C$}}}
\def\gsim{\mathrel{\spose{\lower 3pt\hbox{$\mathchar"218$}}
     \raise 2.0pt\hbox{$\mathchar"13E$}}}
\def\simpropto{\mathrel{\spose{\lower 3pt\hbox{$\mathchar"218$}}
     \raise 2.0pt\hbox{$\propto$}}}
\def\beq{\begin{equation}}
\def\eeq{\end{equation}}
\def\barr{\begin{array}}
\def\earr{\end{array}}
\def\ibid{{\sl ibid.}}
\def\rs{\slash \!\!\!\! R}
\def\tr{{\rm tr\,}}
\def\hc{\rm H.c.}
\def\diag{\rm diag}
\def\cala{{\cal A}}
\def\calo{{\cal O}}
\def\calu{{\cal U}}
\def\calv{{\cal V}}
\def\calq{{\cal Q}}
\def\calm{{\cal M}}
\def\cals{{\cal S}}
\def\sinb{\sin \beta}
\def\cosb{\cos \beta}
\def\tanb{\tan \beta}
\def\barv{\overline{v}}
\def\barm{\overline{m}}
\def\psibar{\overline{\psi}}
\def\vev#1{{\langle#1\rangle}}
\def\etal{ {\it et al.}}
\def\ie{ {\it i.e.} }
\def\eg{ {\it e.g.} }
\def\AJ#1#2#3{{\sl Astr. J.} {\bf #1}, #2 (#3)}
\def\ZPC#1#2#3{{\sl Z.~Phys.} {\bf C#1}, #2 (#3)}
\def\PTP#1#2#3{{\sl Prog. Theor. Phys.} {\bf #1}, #2 (#3)}
\def\PRL#1#2#3{{\sl Phys. Rev. Lett.} {\bf #1}, #2 (#3)}
\def\PRD#1#2#3{{\sl Phys. Rev.} {\bf D#1}, #2 (#3)}
\def\PLB#1#2#3{{\sl Phys. Lett.} {\bf B#1}, #2 (#3)}
\def\PREP#1#2#3{{\sl Phys. Rep.} {\bf #1}, #2 (#3)}
\def\NPB#1#2#3{{\sl Nucl. Phys.} {\bf B#1}, #2 (#3)}
\def\NPBsup#1#2#3{{\sl Nucl. Phys.} {\bf B#1} (Proc. Suppl.), #2 (#3)}
\def\smgaugegroup{{\rm SU(3)_c \otimes SU(2)_L \otimes U(1)_Y}}
\def\tev{{\rm TeV }}
\def\gev{{\rm GeV }}
\def\Mev{{\rm MeV }}
\def\kev{{\rm keV }}
\def\ev{{\rm eV }}
\def\mev{{\rm meV }}
\def\msusy{M_{\rm SUSY}}
\def\mgut{M_{\rm GUT}}
\def\mugut{\mu_{\rm GUT}}
\def\mzino{m_{\tilde {\rm z}}}
\def\mz{m_{\rm z}}
\def\mw{m_{\rm w}}
\def\xiw{\xi_{\rm w}}
\def\xiz{\xi_{\rm z}}
\def\mhl{m_{h^0}}
\def\mha{m_{A^0}}
\def\abs#1{\left|#1\right|}
\def\threebar{\overline{3}}
\def\five{{\bf 5}}
\def\fivebar{\overline{\bf 5}}
\def\ten{{\bf 10}}
\def\tenbar{\overline{\bf 10}}
\def\fivteenbar{\overline{\bf 15}}
\def\xbar{\overline{x}}
\def\drbar{\overline{\rm DR}}
\def\tanb{\tan\beta}
\def\fourth{{1\over 4}}
\def\threetenths{{3\over 10}}
\def\half{{1\over 2}}
\def\threehalf{{3\over 2}}

\def\ifmath#1{\relax\ifmmode #1\else $#1$\fi}
\def\sw{s_{\rm w}}
\def\cw{c_{\rm w}}
\def\sb{s_{\beta}}
\def\cb{c_{\beta}}
\def\tev{{\rm TeV }}
\def\gev{{\rm GeV }}
\def\msusy{M_{\rm SUSY}}
\def\mgut{M_{\rm GUT}}
\def\mhl{m_{h^0}}
\def\mha{m_{A^0}}
\def\abs#1{\left|#1\right|}
\def\tanb{\tan\beta}
\def\fourth{{1\over 4}}
\def\sixteenthirds{\ifmath{{\textstyle{16 \over 3}}}}
\def\Nbar{{\bar N}}
\def\Hbar{{\bar H}}
\def\Tr{{\rm Tr}}
\def\d{{\rm d}}
\def\refmark#1{ [#1]}
\def\alphaem{\alpha_{\rm em}}
\def\tildev{\widetilde V}
\def\mpl{M_{\rm P}}
\def\mweak{m_{\rm weak}}
\def\threebar{{\overline{3}}}
\def\third{\ifmath{{\textstyle{1 \over 3}}}}
\def\sixteenthirds{\ifmath{{\textstyle{16 \over 3}}}}
\def\sevenninths{\ifmath{{\textstyle{7 \over 9}}}}
\def\half{\ifmath{{\textstyle{1 \over 2}}}}
\def\twothirds{\ifmath{{\textstyle{2 \over 3}}}}
\def\fourthirds{\ifmath{{\textstyle{4 \over 3}}}}
\def\centre#1{{\phantom{.}#1\phantom{.}}}
\def\noteadded{$\underline{\hbox{Note Added:}}$}
\def\wr{W_{\slash \!\!\!\! R}}
\def\ET{{\slash \!\!\!\! E_T}}

%
%
\setlength{\parindent}{10pt}
\setlength{\parskip}{.3cm}
\setlength{\textwidth}{16cm}
\setlength{\textheight}{23cm}
\setlength{\topmargin}{-10mm}
\hyphenation{Brems-strah-lung}

\sectionA{1 Introduction}

Theories in which the effects of supersymmetry (SUSY)
breaking are introduced into the ``visible" sector
via the standard model (SM) gauge interactions \cite{original gmsb}
have recently received considerable
attention \cite{new gmsb,gmsb-spectroscopy,%
mu-problem,gravitino-dm,hitoshi,messenger-dm}. One of
the most attractive features of such models is 
the natural explanation for the
smallness of the SUSY-contributions to flavor-changing neutral
current (FCNC) phenomena both in the quark and lepton sector
as a result of a  strongly
constrained sparticle spectrum \cite{gmsb-spectroscopy}.
A complete model of gauge mediated supersymmetry breaking 
requires three sectors:
the ``visible" sector containing the minimal supersymmetric
standard model (MSSM); the ``secluded" 
sector\footnote{%
This sector replaces the so-called ``hidden" sector in
conventional supergravity (SUGRA) models.
The fundamental difference from SUGRA models
is the existence of renormalizable interaction in
addition to gravity and, hence,
a much larger ratio of the visible sector SUSY breaking scale 
over the secluded sector SUSY breaking scale.}
responsible for SUSY breaking
and the messenger sector responsible for the communication of 
SUSY breaking effects into the visible sector.\footnote{%
Recently, there have also been attempts to combine
messenger sector and secluded sector \cite{direct-mess}.}
For the purpose of low energy phenomenology there is no need to 
specify any details about the secluded sector which 
in the minimal version
provides only a
single parameter ($F$) as explained below.
The messenger sector is phenomenologically more important
since it provides all soft SUSY breaking parameters for the MSSM.
One generally assumes, that the messenger sector contains
only complete SU(5) multiplets (\ie pairs of $\five$s and $\ten$s)
in order to naturally maintain
gauge coupling unification \cite{x-unification} ($\tau$-bottom
unification on the other hand is problematic).
The absence of a Landau-pole below the GUT scale limits the number of
additional representations such that $N \equiv N_5 + 3 N_{10}
\lsim 5$ (this value depends somewhat on the messenger masses
assumed to be universal and denoted by $M$).
The general form of the messenger superpotential is given by
\beqn
W = \sum_{\Phi} \lambda^{}_{\Phi} S \bar \Phi \Phi \,,
\eeqn
where $\Phi \subset \five, \ten$ denotes the messenger superfields,
$S$ is an MSSM singlet
belonging to the secluded sector.
In order to suppress FCNC effects we have to forbid any
renormalizable interactions between messenger fields 
and MSSM matter fields by means of a symmetry. 
Hence, all the soft SUSY breaking MSSM parameters are determined
by only two free parameters: $M\equiv \lambda^{}_{\Phi} \vev{S}$ 
and $\Lambda = F/\vev{S}$
with $F \equiv \vev{\partial W^{}_S /\partial S}$
(where $W_S$ is the superpotential of the secluded sector),
and a discrete choice for sign$(\mu)$ and $N$.
From here all the low energy parameters are obtained by renormalization
group evolution \cite{gmsb-spectroscopy}.
In addition, there is the Higgs mass parameter $\abs{\mu}$
determined in the standard fashion by
imposing radiative electroweak symmetry breaking.
The $\mu$ parameter is not a soft SUSY breaking parameter
and a mechanism has to be introduced in order to give it a
phenomenologically allowed value \cite{mu-problem}.
Such a mechanism typically weakens the predictability
of the model by allowing
the soft SUSY breaking Higgs mixing parameter $B_0$,
or the ratio of the Higgs vacuum expectation values (vev)
$\tanb$, to become a free parameter.

One of the most important features of gauge mediated models is the
fact that the gravitino is very light (its mass decreases with the square
of SUSY breaking scale) while its coupling to the SUSY particle spectrum
is enhanced \cite{light gravitino}.
Consequently, the gravitino as the lightest SUSY particle
can be an interesting warm \cite{light gravitino}
and mixed \cite{gravitino-dm} DM candidate, and
the lightest neutralino, often heavier than the gravitino, 
is unstable and no longer a cold dark matter (CDM) 
candidate \cite{neutralino,prexth}.
The goal of this paper is to look for an alternative CDM candidate in the
gauge mediated models.
In section~2, we first reanalyze the properties of 
a messenger sneutrino as a CDM candidate.
We then discuss the feasibility for its direct and indirect
detections in section~3.
Our conclusions are presented in section~4.
Calculations and formul\ae\ are relegated to three appendices.

\sectionB{2 Messenger Sneutrinos as Cold Dark Matter}

Without an interaction of MSSM and messenger sector
in the superpotential, the lightest messenger particle is
stable and a possible CDM candidate.
Several such candidates were investigated in Ref.~\cite{messenger-dm}.
In this paper, we will focus on the electrically neutral components
of chiral supermultiplets, $\five$ and $\fivebar$
which was found to be the most promising possibility \cite{messenger-dm}.
It carries the same gauge quantum numbers as the MSSM scalar
neutrinos and will be referred as messenger sneutrinos.
We denote the electroweak doublets contained in
$\five$ and $\fivebar$ by $\Phi$ and $\bar \Phi$, 
respectively, and define 
$\Phi_\pm = (\Phi \pm \bar \Phi^{\dagger})/\sqrt{2}$.
In this basis the $D$-terms for $\Phi$ and $\bar \Phi$ are
\beqn
D^a      &=& g       \left(\Phi_+^\dagger T^a \Phi_- 
                          +\Phi_-^\dagger T^a \Phi_+\right)\,,\\
D^\prime &=& {g^\prime \over 2}\left(\Phi_+^\dagger  \Phi_- 
                          +\Phi_-^\dagger  \Phi_+\right)\,,
\label{d-term}
\eeqn
and the mass matrix can be written as
\beqn
V = (\Phi_+^\dagger, \Phi_-^\dagger)
\left(\matrix{
 M^2+F & g T^a \vev{D^a}+{g^\prime\over 2} \vev{D^\prime}\cr
g T^a \vev{D_a}+{g^\prime\over 2} \vev{D^\prime}&  M^2-F
}\right)
\left(\matrix{\Phi_+\cr \Phi_-}\right)
\,,\label{phi-matrix}
\eeqn
where $M$ is the messenger mass scale. 
We have assumed that $F$ is real
and the electroweak indices have been suppressed.
Without loss of generality we set $F\geq 0$. 

After electroweak symmetry breaking the $D$ terms
acquire a non-zero vev
that lifts the mass degeneracy of the neutral and charged components
of $\Phi_\pm$. The tree-level masses of the lighter mass eigenstates 
$\phi^Q$ ($Q = 0, -1$ is the electric charge index)
are \cite{messenger-dm}
\beqn
m_{ \phi^Q}^2 =    M^2 -
\sqrt{F^2+(T^3 -Q \sin^2 \theta_W)^2 \mz^4 \cos^2 2\beta}\,.
\eeqn
In the SUSY limit (\ie $F = 0$), the mass squared splitting of 
the neutral ($Q=0$) and the charged ($Q=-1$) components is
of order the $Z$ mass squared $\mz^2$. 
With SUSY-breaking and
in the limit $F\gg \mz^2$, we see that
\beqn
m_{ \phi^-} -m_{ \phi^0}
= { \mz^4 \over 16 F M}\sin^2 2\theta_W \cos^2 2\beta  > 0\,.
\label{tree-level-dm}
\eeqn
This result looks promising: the lightest messenger particle
is neutral. Furthermore, in a large part of the
interesting parameter space the decay
$ \phi^- \rightarrow  \phi^0 f \bar f^\prime$
is sufficiently fast
in order not to disturb big bang nucleosynthesis
(BBN). Unfortunately, as pointed out in Ref.~\cite{messenger-dm},
this $\mz^4$ suppression does in general not apply to 
the one-loop radiative corrections due to
the SUSY breaking $F$-term.
These corrections were found to be dominant and have 
the wrong sign over most of the parameter space,
rendering the lightest messenger sneutrino $\phi^0$
as a CDM candidate questionable.
However, we find that there are large one-loop 
corrections due to electroweak interactions even in the 
limit of unbroken SUSY. Numerically, we find that 
to a good approximation (obtained in the limit 
$M \gg \mz$, $F=0$ and $\tanb = 1$),
\beqn
m_{ \phi^-} -m_{ \phi^0} \simeq {\alpha \over 2} \mz
\simeq 0.3~\gev\,.
\label{deltam}
\eeqn
This result dominates over the tree-level 
term of Eq.~(\ref{tree-level-dm})
and is only slightly modified by one-loop SUSY breaking effects
for $F \ne 0$. Consequently,  the $\phi^-$ 
decay occurs long before BBN and poses no significant constraint
on the parameter space.
The detailed results for the various one-loop diagram calculations
are given in Appendix~A.

We now turn to the CDM relic density obtained from a
freeze-out calculation of the messenger sneutrinos.
The relic density of any particle is governed by
the thermal average of the mean annihilation
rate times the relative velocity,
$\vev{\bar \sigma v}$ at the freeze-out temperature, $T_f$.
Typically, one finds that $T_f \simeq m_{\phi^0}/20\gg 
m_{ \phi^-} -m_{ \phi^0}$
which implies that both charged and neutral components 
are present in the thermal bath at $T_f$.
Hence, the freeze-out calculation should be most properly performed
in the limit of symmetric phase for the four components
($\phi^Q, \phi^{Q\dagger}$).
We also ignore the gauge boson masses, the error 
in this approximation is of order $\mz^2/M^2$ and 
vanishes in the case that $T_f$ is larger than the 
critical temperature of the electroweak phase transition.
We follow Ref.~\cite{swo} for the freeze-out calculation
and our results are summarized in Appendix~B.

In Figure 1, we present our results for the relic density 
($\Omega h^2$) versus
$m_{\phi^0}$ for $r= M/m_{ \phi^0} = 1$ (solid curve) 
and $r = 10$ (dashed curve). The relic density
for $r = 1$ is increased by about 50\% for $r=10$,
and would go up by a factor of 2 for $r\rightarrow \infty$. 
For comparison, we have also plotted the results calculated
in the broken phase \cite{messenger-dm}, by considering
only the lightest messenger sneutrino annihilation.
The rather large discrepancy can be understood as follows:
the broken phase calculation essentially assumes that 
all annihilation cross sections are of the same order,
characterized by $\phi^{0\dagger} \phi^0$ annihilation.
However, notice that
$\sigma( \phi          \phi \rightarrow anything) 
=\sigma( \phi^\dagger  \phi^\dagger
 \rightarrow anything) = 0$;
and $\sigma( \phi^{0 \dagger}  \phi^- \rightarrow anything)$
is suppressed by one power of $x \equiv T_f/m_{ \phi^0}$.
As a result, the relic density calculated in the broken phase
is underestimated by about a factor $4$.
%
%
Generally speaking, as see in Fig.~1,
the sneutrino component of the SU(2) doublet with a mass
in the TeV range could be a good CDM candidate.
Imposing that the universe is not overclosed by the CDM 
particles (\ie\ $\Omega h^2 < 1$)
yields the upper bound on the mass of the lightest messenger
sneutrino
\beqn
m_{ \phi^0} < 3~\tev \,.
\eeqn
For $m_{ \phi^0} \lsim 1~\tev$ it seems
inadequate to neglect Higgs and the squark masses. However,
from Eq.~(\ref{sigmas}) 
we see that the Higgs and matter fields are only produced
in $s$-channel annihilation via intermediate gauge bosons.
Hence, their rates arise at first order in $x$
and only contribute a few percent.

\begin{figure}
\vspace*{13pt}
\vspace*{2.8truein}      
\includegraphics{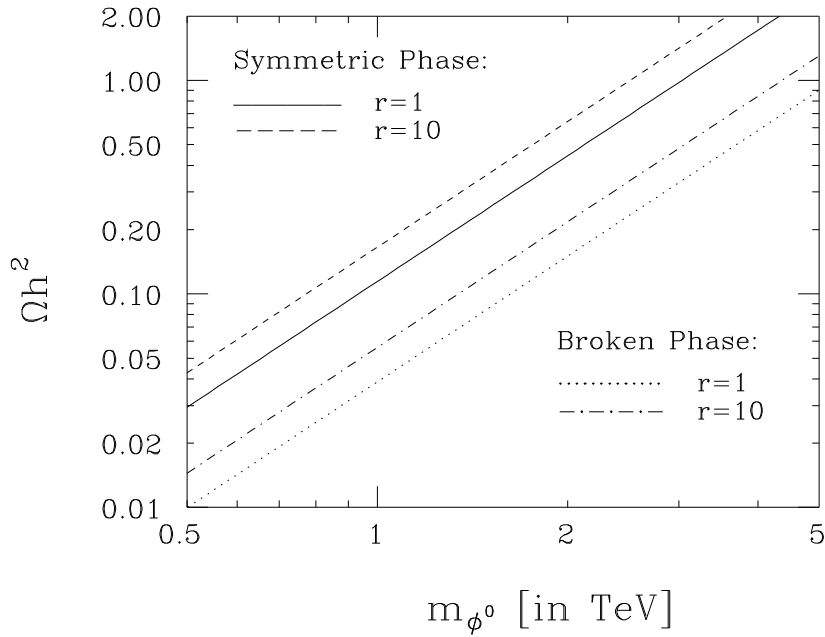}
\caption{The relic density of $\phi^0$ as a function of
$m_{\phi^0}$ in the limit that the mass of all
MSSM particles are much lighter than $m_{ \phi^0}$.
The two upper curves are for our symmetric phase calculation
and the lower ones are for the broken phase.
}
\label{fig1}
\end{figure}

\sectionC{3 CDM Particle Detection}

The most direct way of detecting the existence of dark matter 
particles would be to observe the DM particle-nucleus scattering
by recording the nuclear recoil in a detector \cite{prexth}.
It is known \cite{sneu} that the cross section for the
sneutrino-nucleus scattering 
via the $Z$ exchange is
\beqn
\sigma = { G^2_F \ m^2_{ \phi^0} m_A^2 
\over {2\pi \ (m_{ \phi^0} + m_A)^2} }
[A - 2(1 - 2 \sin^2 \theta_W) Z]^2,
\label{phi-n}
\eeqn
where $G_F$ is the $\mu$ decay constant,
$m_A^{}$ is the nuclei mass, $A$ the atomic mass number 
and $Z$ the atomic number. The current direct searches 
put an upper bound on the cross section. In fact, a scalar
DM particle with mass 3 TeV or less has been ruled out
at a 90\% confidence level \cite{direct-detection}
assuming it accounts for about 30\% or more
of the local galactic halo density 0.3 GeV/cm$^3$.
This would lead to the exclusion of the whole interesting region
of the $m_{\phi^0}$ parameter space in Fig.~\ref{fig1} and
results in an unsatisfactory solution for the CDM issue in the minimal 
version of the gauge mediated model.

To provide a solution to this problem, we propose
to introduce a tree-level interaction 
of the MSSM Higgs sector and messenger sector.
Consider the following superpotential ($W$)
and soft SUSY breaking terms ($V$)
\beqn
W &=& \lambda           N \bar \Phi H_u + {1\over 2}  m_N N^2 \,,\nonumber\\
V &=& \lambda A_\lambda N \bar \Phi H_u + {1\over 2}B_\lambda m_N^{}
N^2 + \hc \,,
\label{poten}
\eeqn
where $N$ is a gauge singlet, $H_u$ the Higgs doublet
coupled to up-type quarks and $ A_\lambda $ and $B_\lambda $ 
the soft SUSY breaking parameters. 
The inclusion of an additional singlet $N$
is motivated by the attempt to understand
the problem of the $\mu$-parameter \cite{mu-problem}, namely,
why the SUSY invariant $\mu$-parameter is of the same order 
as the SUSY breaking scale in the visible sector.
Eq.~(\ref{poten}) may also have an effect on electroweak
symmetry breaking \cite{gian} which deserves further scrutiny.
If both $\lambda$ and $m_N$ are non-zero, then
the messenger number is explicitly broken.
However, the model still respects a $Z_2$ parity
which guarantees the stability of the lightest messenger
particle. 

After EW symmetry breaking, $N$ is no longer a mass eigenstate
and mixes with the neutral component of $\bar \Phi$.
If we assume that  $m_N^{} > M$ and 
$\lambda^2 \vev{H_u}^2 \ll m_N^2 - M^2$, then the mixing is
small so that our analysis in Sec.~2 is still valid. On the
other hand, this mixing has another 
phenomenologically important consequence:
it leads to an additional mass splitting of 
the CP-even and CP-odd components of
the neutral fields
%
\beqn
m_{\rm CP-even}^{} - m^{}_{\rm CP-odd} \ \simeq \
{m_N^{} - M \over 2 m_{\phi^0}} \lambda^2 \vev{H_u}^2
\left[
B_\lambda m_N^{}\ {m_N^{} - M\over (m_N^2 - m_{\phi^0}^2)^2 }
-  {2 A_\lambda \over ( m_N^2 - m_{\phi^0}^2 )}\right]
\label{dmass}\,.
\eeqn
BBN constraints are satisfied if this mass difference 
is larger than a few MeV. For instance, for $M, m_{\phi^0}=1$ TeV,
$m_N^{}=3$ TeV and $B_\lambda,A_\lambda=100$ GeV,
Eq.~(\ref{dmass}) implies $\lambda \gsim {\cal O}(0.1)$.
Since the DM particle coupling 
to the $Z$ boson requires a CP-even and 
CP-odd transition, such a large mass difference would 
prevent DM particle-nuclei scattering via single $Z$ exchange 
from happening. This is because the initial kinetic energy in
the c.~m.~frame
for DM particle-nuclei scattering is typically of order 0.1 MeV 
(assuming the DM particle velocity is about $10^{-3}c$)
and is much smaller than the CP-even and CP-odd mass difference.
The dominant contribution to DM particle-nuclei 
scattering would therefore be the spin-independent Higgs exchange
through quarks and quark loops. In analogy to the neutralino-nucleon
scattering \cite{drees-nojiri}, we find  that the cross section
for sneutrino-nucleus scattering 
via Higgs exchanges is
\beqn
\sigma_{\rm nucl} = {m_A^2 \over {4\pi \
(m_{ \phi^0} + m_A)^2 } }
\ [f_p Z + f_n (A-Z)]^2 ,
\label{phi-n-higgs}
\eeqn
where $f_p \simeq f_n$ are the effective sneutrino-nuclei coupling
via Higgs exchanges. They are given in Apendix C.
Following the discussion in Ref.~\cite{drees-nojiri}, 
we estimate the  direct search event rate ($R_D$) 
for scalar neutrino-nuclei scattering by 
\beqn
R_D \approx {1.8\times 10^{11} {\rm GeV^4} \over {\rm kg \ day} } \
{\zeta \ \sigma_{\rm nucl} \over m_{ \phi^0} m_A} ,
\label{direct}
\eeqn
where $\zeta$ describes the suppression due to 
the nuclear form factor \cite{form}. 

Another way of detecting the DM particle is by the large scale neutrino
detectors via annihilation products from DM particles inside the
sun/earth. Following the arguments in Ref.~\cite{fred}, 
the expected event rate in a neutrino detector 
induced by $\nu_\mu$ from the $ \phi^{0\dagger} \phi^0$ 
annihilation in the sun may be estimated  by  
\beqn
R_{ID} \approx {2.65\times 10^{39} \over {10^4\ {\rm m^2 \ yr}} }\
{ m_{\phi^0} \over {\rm GeV}} \ \eta \
\sum_A \kappa_A^{} \ \zeta \ {\sigma_{\rm nucl} \over {\rm cm^2}} ,
\label{indirect}
\eeqn
where $\kappa_A^{}$ is the relative abundance of element $A$
with respect to hydrogen in the sun. With the nuclear form factor
$\zeta$ included, we find that the leading contributions to the
DM particle capture are from the elements iron, oxygen and helium.
$\eta$ in Eq.~(\ref{indirect}) is the neutrino
escape probability in the sun parameterized by \cite{fred}
\beqn
\eta = \left[1 + {m_{\phi^0} \over 5.2~\tev}\right]^{-7},
\label{suppress}
\eeqn
which significantly suppresses the $\nu_\mu$ flux for 
$m_{\phi^0}\ge 1$ TeV. The background rate from the flux of
atmospheric neutrinos coming from a pixel around the sun
is estimated to be \cite{fred}
\beqn
B_{ID} \approx {0.11 \over {10^4\ \rm m^2 \ yr}}\
\left( {m_{ \phi^0} \over \tev}\right)^{-2} .
\label{Bindirect}
\eeqn
\begin{figure}
\vspace*{13pt}
\vspace*{3.8truein}      
\includegraphics{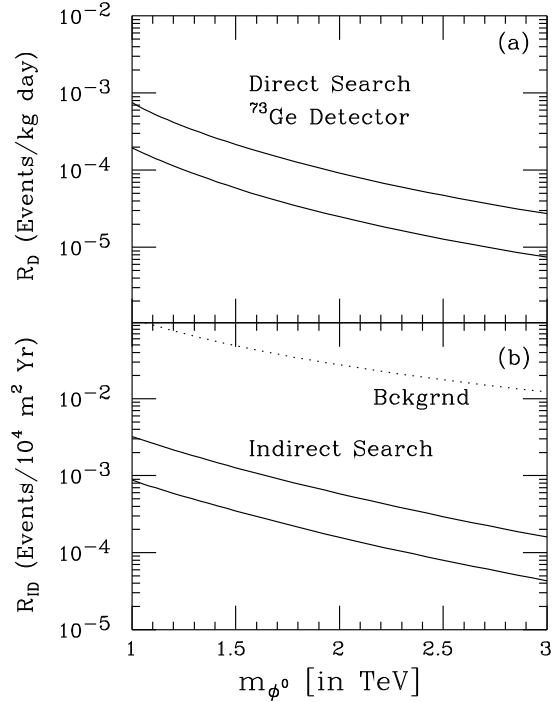}
\caption{The event rate for DM particles
consisting of messenger sneutrinos in direct and indirect searches
as a function of their mass, $m_{\phi^0}$.
}
\label{fig2}
\end{figure}

The scattering cross section 
given by Eq.~(\ref{phi-n-higgs}) is rather small.
For a 1 TeV scalar neutrino scattering off a nucleon,
it is typically of order $10^{-47}\sim 10^{-44}$ cm$^2$,
depending on SUSY parameters.
Comparing the direct and indirect detection rates
given in Eqs.~(\ref{direct}) and (\ref{indirect}),
we see that the indirect search seems to be more
suitable for a heavier DM particle. Especially,
the background decreases like $1/m^2_{ \phi^0}$.
However, the suppression due to $\zeta$ and $\eta$
in Eq.~(\ref{suppress}) becomes increasingly 
severe for heavier $\phi^0$. Yet, Eq.~(\ref{indirect})
may have overestimated the signal somewhat due to the omission
of a correction factor to the capture rate by the sun 
for a heavier DM particle \cite{prexth}. 
The signal rates for the direct and indirect searches
are shown in Fig.~2 as a function of $m_{ \phi^0}$. 
The rate for direct search in Fig.~2(a) is for a 
$^{73}$Ge detector. The lower solid curve corresponds
to a high mass limit for the CP-odd Higgs boson 
so that the dominant contribution
is from a SM-like Higgs boson, assuming $m_h=100$ GeV.
The upper solid curve is for an optimal choice of parameters
to enhance the signal rate (CP-odd Higgs mass of 80 GeV 
and $\tanb=50$). In our numerical analysis,
the coupling $\lambda$ in Eq.~(\ref{poten})
has been taken to be unity.
We see that the rate is typically of order $10^{-4}$ 
events/(kg~day). It is much below the current experimental
sensitivity of about 2 events/(kg~day) \cite{direct1},
and is also unreachable by next generation experiments with sensitivity
of $10^{-2}$ events/(kg~day) \cite{prexth}.
Fig.~2(b) shows the calculated signal rate for
the indirect search, with the same sets of SUSY
parameters as in (a). Based on Eq.~(\ref{Bindirect}),
the background rate is calculated and shown by the dotted curve in 
Fig.~2(b). Again, it is difficult to detect the signal for the
indirect search as well. 

\sectionD{4 Conclusions}

We have considered the possibility that the CDM constitutes of 
sneutrino-like messenger particles.
We found that the neutral components $\phi^0$ of the SU(2) doublet 
is naturally lighter than the charged component over 
most of the relevant parameter space due to one-loop
electroweak corrections and could serve as
a CDM candidate.
However, our relic density calculation shows that a 
significant amount of messenger CDM in the minimal model
has been already ruled out by present experimental results.
We introduce a mechanism that generates a CP-even--CP-odd 
mass splitting. This circumvents the constraints 
from present detection experiments and allows the messenger 
sneutrino to be a valid CDM candidate in a wide range of SUSY 
parameter space. Consequently, the
detection via direct and indirect searches
would be very difficult due to the rather small 
messenger sneutrino-nucleon scattering cross section.

\sectionE{5 Acknowledgement}
We would like to thank M. Drees and F. Halzen for illuminating
discussions. This work is supported in part by the US
Department of Energy under grant  DE-FG03-91ER40674, 
and in part by the Davis Institute for High Energy Physics.

\appendixA{Appendix A: 
$m_{ \phi^-}$--$m_{ \phi^0}$ mass difference }

At tree-level, the only source of a mass splitting of 
different members of an
SU(2)$_L\otimes$U(1)$_Y$ are a non-zero vev for the SU(2) D-term
which is strongly suppressed as shown in Eq.~(\ref{tree-level-dm}).
In this appendix we present the result for the dominant one-loop
contribution to this mass difference.
A complete one-loop calculation is rather involved.
In particular, it requires a renormalization of $\tanb$
which enters already at tree-level.
Hence, we consider the case $\tanb = 1$ where the corrections to the
mass squared difference
is finite and given by the difference
of the self-energies
\beqn
\Delta m^2 \equiv
m^2_{ \phi^-} - m^2_{ \phi^0} =
A_{ \phi^-  \phi^-}(m_{ \phi^-}^2) - 
A_{ \phi^0  \phi^0}(m_{ \phi^0}^2) \,.
\eeqn
Furthermore, we neglect terms suppressed by
$F/M^2$ ({\it i.~e.}, we set $F=0$ and as a result
$m_{\phi^0} = m_{ \phi^-} = M$).

There are three types of diagrams that contribute to the 
self-energies: the loops involving gauge bosons ($g$), 
gauginos ($\tilde g$), and
loops with tri-linear terms arising from the $D$-terms (Higgs field only).
Loops with quartic terms arising from the $D$-terms 
do not generate mass splitting among the different
members of the multiplet $ \phi$.
The results are
\beqn
A_{\phi^0 \phi^0}^g (M^2) &=& {\alpha\over 4 \pi \sin^2 2\theta_W}
\left[\left(4  M^2 -\mz^2\right)B_0(M^2, \mz^2, M^2) - A_0(\mz^2) 
- A_0(M^2) \right] \,,
\nonumber \\
A_{\phi^0 \phi^0}^{\tilde g}(M^2) &=& {\alpha\over 2 \pi \sin^2 2\theta_W}
\left[\mzino^2 B_0(M^2, \mzino^2, M^2) + A_0(\mzino^2)
 + A_0(M^2)\right] \,,
\nonumber \\
A_{\phi^0 \phi^0}^{H}(M^2) &=& {-\alpha\over 4 \pi \sin^2 2\theta_W}
\mz^2 B_0(M^2, m_{H^0}^2, M^2) \,,
\eeqn
where  $\alpha$ is the fine structure constant and
$\theta_{W}$ the weak mixing angle.
In the SUSY limit we have $m_{H^0} = \mzino = \mz$.
The self-energy for the charged component can be written as
\beqn
A_{\phi^- \phi^-} (M^2) =
  \cos^2 2 \theta_W A_{\phi^0 \phi^0} (M^2)
+ \sin^2 2 \theta_W A_{\phi^0 \phi^0} (M^2)(\mz^2\rightarrow 0)
\eeqn
Hence, we obtain
\beqn
\Delta m^2 = {\alpha\over 4 \pi} M^2 
\left[B_0(M^2, 0, M^2) - B_0(M^2, \mz^2, M^2)\right]
\,.\eeqn
In dimensional regularization,
the scalar one and two point functions are given by
\beqn
16 \pi^2 A_0(m_1^2) &=& m_1^2 \left(\Delta + 1 - \ln m_1^2\right) \,,
\nonumber\\
16 \pi^2 B_0(p^2,m_1^2, m_2^2) &=& \Delta
- \int_0^1 \d x \ln \left[x m_1^2 + (1-x)m_2^2 - x(1-x) p^2\right]\,,
\eeqn
where $\Delta = 1/\epsilon - \gamma_E^{} + \ln 4\pi$ 
in $d=4-2\epsilon$ dimensions, with $\gamma_E^{}$ the Euler constant.
Expanding Eq.~({A.4}) for $M \gg \mz$, Eq.~(\ref{deltam}) is
recovered.

\appendixB{Appendix B: Annihilation Rate and Relic Density}

Our calculation for the annihilation rate 
follows the formalism developed in  Ref.~\cite{swo}.
There are four nearly mass-degenerate particles present
at the freeze-out temperature $T_f$, 
$\phi^Q$ and $\phi^{Q\dagger}$ ($Q = 0, -1$).
If we assume that there is no messenger number asymmetry
then the total relic density is
$n = 4 n_{\phi^Q} = 4 n_{\phi^{Q\dagger}}$
and the rate equation can then be written as
\beqn
{\d n\over \d t} &=&  - 3 H n - \vev{{\overline \sigma} v_{\rm rel}} n^2 \,,
\nonumber\\
\eeqn
where $H$ is the expansion rate of the universe and
\beqn
{\overline \sigma}&\equiv & {1\over 8} \sum_{Q, Q^\prime}
 \sigma_{\phi^{Q\dagger} \phi^{Q\prime}} \, .
\eeqn
Note $ \sigma_{\phi^Q          \phi^{Q^\prime}} 
      = \sigma_{\phi^{Q\dagger} \phi^{Q^\prime\dagger}} =0$.

Following Ref.~\cite{swo}, we first define
a Lorentz invariant function
\beqn
w(s) \equiv {\beta \over 64 \pi} \int_{-1}^{1}
{1\over 8}\sum_{Q, Q^\prime, a, b}
\abs{{ {\cal M}(\phi^{Q\dagger} \phi^{Q^\prime} \rightarrow a b)}}^2
 \d \cos \theta
\eeqn
where $\beta^2 = 1- 4 m_{\phi}^2/s$ is the squared velocity
and the sum over $a$ and $b$ symbolizes the sum
over all two-body final states. Then
\beqn
w(s) = \sum_{g, g^\prime = b, w}
\left(w^{g g^\prime} + w^{\tilde g \tilde g^\prime}\right)
+w^{ s^\dagger s}
+w^{\bar f f}\,,
\eeqn
the result of the expansion of the different contributions
to $w(s)$ to first order in $\beta^2$ is
\beqn
w^{ww} &=& {g^4\over 512 \pi}
   \left(3 - 2 \beta^2\right)\,,
\nonumber\\
w^{bw} &=& {g^2 g^{\prime 2} \over 512 \pi}
   \left(3 - 4\beta^2\right)\,,
\nonumber\\
w^{bb} &=& {g^{\prime 4}\over 512 \pi }
\left(1 - {4\over 3} \beta^2\right) \,,
\nonumber\\
w^{\tilde w \tilde w} &=& {g^4\over 512 \pi}
    \left[{12 r\over (1+r)^2}
+{4 \beta^2 \over (1+r)^4}
\left(1 + 7 r - 8 r^2 + 3 r^3 +r^4 \right)\right]\,,
\nonumber\\
w^{\tilde b\tilde w} &=& {g^2 g^{\prime 2} \over 512 \pi}
\left[{12 r\over (1+r)^2} - {4 r \beta^2 \over (1+r)^4}
\left( 1 + 6 r- 3 r^2\right)
\right]\,,  
\nonumber\\
w^{\tilde b\tilde b} &=& {g^{\prime 4} \over 512 \pi}
\left[{4 r\over (1+r)^2} - {4 r \beta^2
\over (1+r)^4}\left({1\over 3} + 2r-r^2\right)
\right]\,,
\nonumber\\
w^{ s^\dagger s} &=& \sum_s {N_c \beta^2
\over 4096 \pi}
\left[N_{\rm w} Y_s^2 g^{\prime 4} + 6 \left(N_{\rm w} - 1\right) g^4 
\right]
\,,
\nonumber\\
w^{\bar f f} &=& 2 w^{ s^\dagger s}
\,, 
\label{sigmas}\eeqn
where $r = M/m_{\phi^0}$. The result for the gauge bosons 
(gauginos) corresponding U(1)$_Y$ and SU(2)$_L$ are denoted by 
a superscript $b$ and $w$ ($\tilde b$ and $\tilde w$), respectively.
Here, $N_c = 1$ ($3$) for colored   singlets (triplets)
  and $N_{\rm w} = 1$ ($2$) for SU(2)$_L$ singlets (doublets).
The index $s\ (f)$ runs over all MSSM scalar (fermion) fields
other than gauge bosons and gauginos.

The thermal average $\vev{{\overline \sigma} v_{\rm rel}}$ 
can be expressed by 
\beqn
\vev{{\overline \sigma} v_{\rm rel}} =
{1\over m_{ \phi^0}^2} (A + B x) \,,
\eeqn
where $x=T_f/m_{\phi^0}$, and \cite{swo}
\beqn
A = w(4 m_{\phi^0}^2) , \quad 
B = 3 \left[2 m_{ \phi^0}^2\ {\d w(s)\over \d s} - 
w(s)\right]_{s = 4 m_{ \phi^0}^2}\,.
\label{thermal}
\eeqn
%
We have ignored the possible contribution to the
annihilation cross sections from the interactions in
Eq.~(\ref{poten}). This implies 
a lower limit of $\vev{{\overline \sigma} v_{\rm rel}}$
and hence, an upper limit on the relic density.

It is customary to express the relic abundance in terms of 
the mass density in units of the critical density 
$\Omega_{\phi^0}=\rho/\rho_c$. It is found that
\beqn
\Omega_{\phi^0} h^2 =  {8.5 \times 10^{-5} \over \sqrt{g_*} }
\ \left({m_{\phi^0}\over {\rm TeV}}\right)^2
\ {x^{-1} \over A+{1\over 2}B x} \ ,
\eeqn
where $h$ is the Hubble constant, $g_*\simeq 228.75$ \cite{messenger-dm}.
The value $x$ is related to the freeze-out temperature 
and can be obtained iteratively by
\beqn
x^{-1} \equiv  m_{\phi^0}/ T_f 
= \ln \left[{0.076\over \sqrt{g_*}} 
{M_{Pl}\over m_{\phi^0}} ( A+B\ x)\ \sqrt{x} \right] ,
\eeqn
where $M_{Pl}$ is the Planck mass. 
%
%

\appendixC{Appendix C: sneutrino-nuclei Elastic Scattering
via Higgs exchange }

For the model of our interest, the $Z$-exchange contribution
between two different sneutrino mass eigenstates is negligible.
The leading contribution would therefore be the spin-independent
Higgs exchange. We write the effective interaction between
a sneutrino ($ \phi $) and  a nucleon ($N$)
\beqn
{\cal{L}}_{\phi N} = f^{}_N \  \phi^\dagger   \phi \
\bar \psi_N \psi_N ,
\eeqn
where $f^{}_N \ (N=p,n)$ is the effective coupling through Higgs boson
exchanges:
\beqn
f^{}_N = m^{}_N \ ( \sum_q^{u,d,s} K_q + 
           \sum_Q^{c,b,t} K_Q + 
           \sum_{\tilde Q}^{{\tilde c},{\tilde b},{\tilde t}} 
                 K_{\tilde Q}) ,
\eeqn
where $K_q$, $K_Q$ and $K_{\tilde Q}$ are contributions from 
a light quark $q$, a heavy quark $Q$ and its supersymmetric 
partner $\tilde Q$. 
In analogy to the neutralino-nucleon interaction \cite{drees-nojiri},
we determine the couplings as
\beqn
K_q =  f_{T_q}^{} \sum_j {c^j_\phi c^j_q\over m^2_{H_j}} ,\quad
K_Q =   {2 \over 27}\ f_{T_G}^{}
\sum_j {c^j_\phi c^j_Q\over m^2_{H_j}} ,\quad
K_{\tilde Q} = {1 \over 108}\ f_{T_G}^{} 
\sum_j {c^j_\phi c^j_{\tilde Q}\over m^2_{H_j} m^2_{\tilde Q}}\, .
\eeqn
The constant $f_{T_q}^{}$ is the nucleon mass fraction due to a
light quark $q$, and $f_{T_G}^{} = 1-\sum_q^{u,d,s} f_{T_q}^{}$.
Numerically, we take \cite{cheng}
\beqn
{\rm for \ a \ proton}: \quad f_{T_u}^{}= 0.023, \ \ f_{T_d}^{}= 0.034 ,\\
{\rm for \ a \ neutron}: \quad f_{T_u}^{}= 0.019,\ \ f_{T_d}^{}= 0.041 ,
\eeqn
and \cite{strange}
\beqn
 f_{T_s}^{}= 0.14 .
\eeqn
The $H_j^0 \phi^\dagger  \phi$ couplings are given by 
\beqn
c^{h^0}_\phi = {1\over 2} \lambda^2 v\ \sin\beta \cos\alpha, \quad
c^{H^0}_\phi = {1\over 2} \lambda^2 v\ \sin\beta \sin\alpha ,
\eeqn
where $\lambda$ is a model-dependent parameter in Eq.~(\ref{poten}), 
$v = 246~\gev$ is the Higgs vacuum expectation value
and $\beta$ and $\alpha$ are the standard mixing parameters in the
SUSY Higgs sector.
Finally, $c^j_{q,Q,{\tilde Q}}$ are the couplings of a Higgs boson
$H_j$ to quarks and squarks and we have followed the convention in
Appendix A of Ref.~\cite{drees-nojiri}. In practice, the squark
contributions are small and have been neglected.

The cross section for the coherent
elastic scattering of $ \phi$ off a nuclei $A$ 
in the non-relativistic limit is thus calculated to be
\beqn
\sigma = 
{ {\bar \Sigma} |{\cal{M}}|^2 \over 16\pi \ (m_{ \phi} + m_A)^2  }
= { m_A^2 \over 4\pi \ (m_{ \phi} + m_A)^2 }
\ [f_p\ Z + f_n\ (A-Z)]^2 .
\eeqn
%


\end{document}